\documentclass[]{aa}
\usepackage{graphicx}
\usepackage{times}
\usepackage{natbib}
\usepackage[below]{placeins}
\usepackage{psfig}

\newcommand{\rxj}{RX~J0720.4-3125}
\newcommand{\rxjb}{RX~J1856.5-3754}

\newcommand{\rosat}{{\em ROSAT}}
\newcommand{\chandra}{{\em Chandra}}

\newcommand\xmm{{\em XMM-Newton}}

\begin{document}

\titlerunning{The X-ray emission of RX J0720.4-3125}
\title{Long-term variability in the X-ray emission of RX~J0720.4-3125}


\author{Cor P. de Vries \inst{1} \and
	Jacco Vink \inst{1} \and
	Mariano M\'{e}ndez \inst{1} \and
	Frank Verbunt \inst{2}
	}

\offprints{C.P. de Vries, \\
	   \email{C.Pde.Vries@sron.nl}}

\institute{SRON, National Institute for Space Research, 
		Sorbonnelaan 2, 3584 CA Utrecht, The Netherlands
	   \and
	   Astronomical Institute, Utrecht University, 
	   PO Box 80000, 3508 TA Utrecht, The Netherlands
	   }

\date{Received $\prec date1 \succ$, accepted  $\prec date2 \succ$}

\abstract{We detect a gradual, long-term change in the shape of the
X-ray spectrum of the isolated neutron star RX J0720.4-3125, such that
the spectrum of the source can no longer be described as a blackbody
spectrum. The change is accompanied by an energy-dependent change in
the pulse profile. If the X-ray emission is influenced by the magnetic
field of the pulsar, these changes in spectral shape may point to
precession of the neutron star.
\keywords{neutron stars, X-rays}
}

\maketitle

\section{Introduction}

\rxj\ \citep{haberl97} belongs to the group of radio-quiet neutron
stars  discovered by \rosat. As these objects are characterised by soft
X-ray emission from the surface, it was hoped that high resolution
spectroscopy of those sources with the \chandra's Low Energy Grating
Spectrometer (LETGS) or \xmm's  Reflection Grating spectrometer (RGS)
would reveal line features from a thin atmosphere that could provide
information on the surface gravity  of the neutron
star and hence on the equation of state of the neutron star matter.
However, deep observations with \xmm\ and \chandra\ of
\rxj\ \citep{paerels01,kaplan03} and \rxjb\ \citep{burwitz01}  have in
that respect been somewhat of a disappointment,  as the spectra are
almost perfectly well described by a blackbody spectrum. The only
unusual feature of the X-ray spectra is that the blackbody emission
implies a radius too small for a neutron star and, moreover, 
underpredicts the observed optical and UV flux
\citep{kaplan03,burwitz01,motch03}. This implies that another cooler
component is present and that the X-ray emission only comes from part
of the surface.

\rxj\ has a low absorbing column density ($N_{\rm{H}} \approx 
10^{20} {\rm cm}^{-2}$), is at an approximate distance of $\approx 300
{\rm pc}$ \citep{kaplan03}, and shows sinusodial pulsations with a
period of 8.39~s \citep{haberl97,cropper01}. The upper limit to the
period derivative $\dot P < n\times10^{-13}$\,ss$^{-1}$, with $n$ `a
few' \citep{kaplan02}, implies a magnetic field strength
$B<3\sqrt{n}10^{13}$\,G and a characteristic age $\tau_c\equiv0.5P/\dot
P>(1.3/n)$\,Myr.  As \rxj\ was considered to be a perfect blackbody
spectrum  with a temperature of $k{\rm T}_{\rm BB} \approx 86~{\rm eV}$
it was observed several times by \xmm\ for calibration purposes.
However, as we will demonstrate here, the spectrum of \rxj\ has slowly
hardened and cannot be described any more by a blackbody spectrum,
while the pulse shape has become narrower and the phase dependence of the
spectrum changed with time.

While we were finalising this paper, \citet{haberl03b} released a
pre-print in which they show that the spectrum of \rxj\ depends on the
pulse phase. They also discuss changes in the spectrum as found with
EPIC, but assign  these to calibration inaccuracies.

\section{Observations}

\begin{table}
	\caption[]{
\xmm\ observations of RX J0720.4-3125}
	\label{obs}
   \begin{tabular}{ccccc}
	Orbit & Observation & Date & Julian date & Duration (ksec) \\
\hline 
	0078 & 0124100101 & 13-05-2000 & 2451677.9 & 65 \\
	0175 & 0132520301 & 21-11-2000 & 2451870.3 & 30 \\
	0533 & 0156960201 & 06-11-2002 & 2452585.4 & 30 \\
	0534 & 0156960401 & 08-11-2002 & 2452498.5 & 32 \\
	0622 & 0158360201 & 02-05-2003 & 2452762.5 & 81 \\
	0711 & 0161960201 & 27-10-2003 & 2452940.5 & 45 \\
   \end{tabular}
\end{table}

Table~\ref{obs} gives the log of \xmm\ observations of \rxj. Since the
source spectrum is rather soft, and most of the emission comes from
below $\sim 1.8$ keV, we use RGS \citep{herder} data for the spectral
analysis; due to the nature of the gratings, the effective area of the
RGS is rather insensitive to changes in CCD gain and charge transfer
inefficiency. The RGS was operated in normal spectroscopic mode,
yielding a time resolution of 4 s, making it unsuitable for a proper
timing analysis. Therefore, for this purpose we use EPIC/PN
\citep{struder01} since it has the largest effective area among the
instruments on board \xmm, and operates in a high-time resolution mode.

During the first four observations, PN was operated in full-frame mode,
which produces an image of 378 $\times$ 384 pixels with a time
resolution of 73.4 ms. During the last two observations the PN was
switched to small-window mode, yielding images of 63 $\times$ 64 pixels
with a time resolution of 6 ms. The thin filter was used in revolutions
0078, 0533, and 0534, the medium one in revolution 0175, and the thick
one in revolution 0622. The observation of revolution 0711 started with
the thin filter, and later switched to the medium one. All data were
processed using the standard \xmm\ software SAS version 5.4.1.


\subsection{Spectral Analysis}

\begin{figure}
  \resizebox{\hsize}{!}{\includegraphics[angle=0,clip]{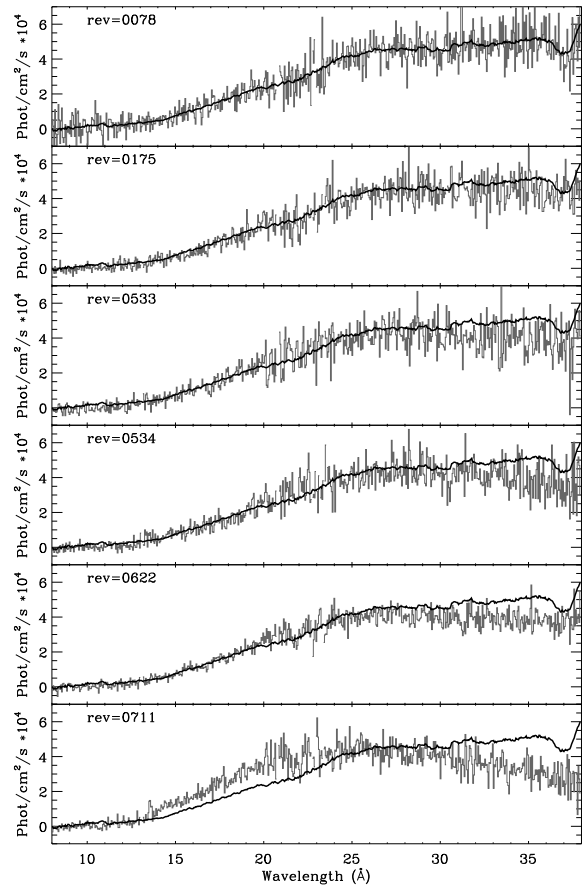}}
  \caption{ Fluxed RGS spectra of the different observations of \rxj. 
    The spectra are shown as the light grey histograms. 
   The black lines are the smoothed data from orbit 0078. 
    The differences between the lines and the different data show the
    variations of the spectrum with respect to the
    data of orbit 0078. Especially the large change at orbit 0711 with an 
    increase of flux between 15 and 25~\AA~can be clearly recognized. } 
  \label{fluxed}
\end{figure}

Figure~\ref{fluxed} shows the RGS spectra for the 6 revolutions. 
The indivual RGS's agree with each other within the statistical uncertainties,
so the data from both RGS's have been combined. From
the Figure it is apparent that the spectrum of \rxj\ gradually
changes, while the total flux appears to remain fairly constant. The
change of the spectrum is most noticeable in orbit 0711, where there is
an increase of the flux in the 15 \AA\ to 25 \AA\ range, and a drop in
the flux above 30 \AA.

To verify that the RGS instrument is stable and that its effective area 
does not change, we have also analyzed data from two sources that are
stable on a time scale of years. 

The first source is the supernova remnant 1E 0102-72.3 in the Small
Magellanic Cloud, a calibration source that is regularly observed with
the RGS. From the strength of the bright emission lines from this
source in the range 13 to 23~\AA, as measured in data from orbits 0065
to 0711, we find that the effective area of the instrument in this
range changes by less than 3\%. In the long wavelength range the SNR
0102-72.3 spectrum has a weak carbon line at 33.7~\AA. The upper limit
on possible changes in the effective area of the RGS instrument at this
wavelength is about 10\%, mainly determined by the limited statistics
in this line.  All data from 1E 0102-72.3 are consistent with no change
in the RGS effective area. The second source is the calibration source
Mrk~421. An upper limit to any change in the neutral Oxygen absorption
edge around 22.8~\AA, which is a tracer of possible instrument contamination (e.g.
ice), is 5\%.

Since all upper limits on possible changes in effective area are well
below the changes seen in \rxj\, we conclude that the X-ray spectrum of
\rxj\ itself is subject to change over the course of about 3 years.

For the spectral analysis we use the {\em XSPEC}  spectral fitting
package \citep{xspec}. As noted by \citet{paerels01} the data of orbit
0078 can be well fitted by an absorbed blackbody spectrum. With the
improved calibration of the RGS we find a temperature of $kT_{\rm BB} =
86.7\pm0.3$~eV and an interstellar  column density of $N_{\rm H} =
(1.41\pm0.07)\times10^{20}$~cm$^{-2}$, in reasonable agreement with the
values obtained with the \chandra\ LETGS instrument, $kT_{\rm BB} =
81.4\pm1.3$~eV and  $N_{\rm H} = (1.32\pm0.14)\times10^{20}$~cm$^{-2}$,
by \citet{kaplan03}. The LETGS observation was made in February 2000,
four months before the first \xmm\ observation. 

Fits with the same model to the remaining spectra yield a temperature
increase with time, as well as changes as large as $\Delta N_{\rm H} \sim
5\times10^{20}$~cm$^{-2}$ in the interstellar absorption. The model,
however, does not provide satisfactory fits to all the spectra.

Next, we fit all the data simultaneously. Since large variations of the
interstellar absorption over such a short period are unlikely, we
assume that $N_{\rm H}$ remains constant; we therefore couple this
parameter for all observations, while we let the blackbody temperature and 
normalization vary freely between observations
(Table~\ref{temp}). Although these fits serve to emphasize the gradual
hardening of the spectra, half of the spectra, especially that of
revolution 0711, are poorly fit by this model.

Following \citet{haberl03a} and \citet{vankerkwijk03}, next we fit the
data with a model that consists of a blackbody affected by additional
absorption by a broad gaussian line, as might be expected in the case
of cyclotron absorption. In these fits, we constrain the parameters of
the blackbody model to be the same for all the observations, whereas the
parameters of the line are allowed to vary independently. Although
this model fits the data well, in most cases we find that the width of
the line is larger than its central energy, or the central energy of
the line lies outside the spectral range covered by the data. In those
cases, the tail of the broad gaussian just serves to attenuate the
emission of the source at long wavelengths. Furthermore, we find no
clear trend in the central energy or the width of the line.

We find that the ratio of the spectra of the last and the first
observation is close to a power law. We therefore fit
the data to an empirical model that consists of a blackbody multiplied
by a power law $E^{\Gamma}$, all affected by interstellar absorption. While the index
of the multiplicative power law is allowed to change between
observations, for these fits we constrain the parameters of the
blackbody and the interstellar absorption to be the same in all
observations. While it is difficult to assign a physical interpretation
to this model, it provides an acceptable description of the data in the RGS
range ( 10 - 38~\AA, see Table
2), it has fewer parameters than the gaussian absorption model and, in
addition, the  index of the power law increases steadily over the
course of the observations.

{\scriptsize
\begin{table*}
        \caption[]{Spectral fits$^a$ to the observations of RX J0720.4-3125}
        \label{temp}
   \begin{tabular}{ccccccc}
Orbit & $kT$ (eV) & $L_{\rm bb}^b$  & $\Gamma^c$ & \multicolumn{3}{c}{Flux $^d$}           \\
      &           &                 &            & (10-23 \AA) & (23-38 \AA) & (10-38 \AA) \\
\hline
0078  & 81.3(3)   & 2.54(3)         &  1.44(5)   & 1.86(4)     & 4.72(7)     & 6.58(10)    \\
0175  & 81.4(5)   & 2.42(3)         &  1.45(5)   & 1.77(5)     & 4.49(7)     & 6.26(11)    \\
0533  & 84.6(4)   & 2.22(3)         &  1.74(5)   & 1.89(4)     & 4.20(7)     & 6.09(11)    \\
0534  & 84.0(4)   & 2.28(3)         &  1.67(5)   & 1.88(4)     & 4.31(7)     & 6.19(11)    \\
0622  & 86.3(3)   & 2.07(2)         &  1.85(6)   & 1.93(4)     & 4.08(7)     & 6.01(06)    \\
0711  & 98.2(4)   & 1.96(2)         &  2.70(8)   & 2.81(4)     & 4.01(6)     & 6.82(09)    \\
   \end{tabular}
{\scriptsize
\begin{list}{}{}
 \item[$^a$] Fit range 10--38~\AA. For the interstellar absorption
             component we use cross sections from \citet{verner96},
             and abundances from \citet{wilms00}. Numbers in
             parentheses are 1-$\sigma$ confidence limits in the last
             digit(s). For the fits with a blackbody, the best-fit
             value for $N_{\rm H}$, constrained to be the same in all
             observations, is $(4.3 \pm 0.1) \times 10^{20} {\rm
             cm}^{-2}$. For the fits with a blackbody times a power
             law, the best-fit values for $N_{\rm H}$, $kT$, and the
             bolometric luminosity of the blackbody, constrained to be
             the same in all observations are, respectively, $(1.28
             \pm 0.1) \times 10^{20} {\rm cm}^{-2}$, $70.3 \pm 0.1$
             eV,  and $(3.04 \pm 0.02) \times 10^{32} d^2_{\rm 300}$
             erg s$^{-1}$, with $d_{\rm 300}$ the distance to \rxj\ in
             units of 300 pc. For the fits with a blackbody model,
             $\chi^2 = 874.1$ for $662$  degrees of freedom. For the
             fits with a blackbody times a power-law model $\chi^2 =
             776.8$ for $668$ degrees of freedom.
 \item[$^b$] Bolometric luminosity of the blackbody in units of
             10$^{32} d^2_{\rm 300}$ erg s$^{-1}$, where $d_{\rm 300}$ is
	     the distance to the source in units of 300 pc.
 \item[$^c$] Index of the power law in the model that consists of a
             blackbody times a power law (see text)
 \item[$^d$] Observed flux, obtained from the fits with a blackbody
             times a power-law; only statistical errors are indicated. The systematic
             errors in the flux are of order 5--10\%.
\end{list}
}
\end{table*}

}

\subsection{Timing analysis}

\begin{figure}
\psfig{figure=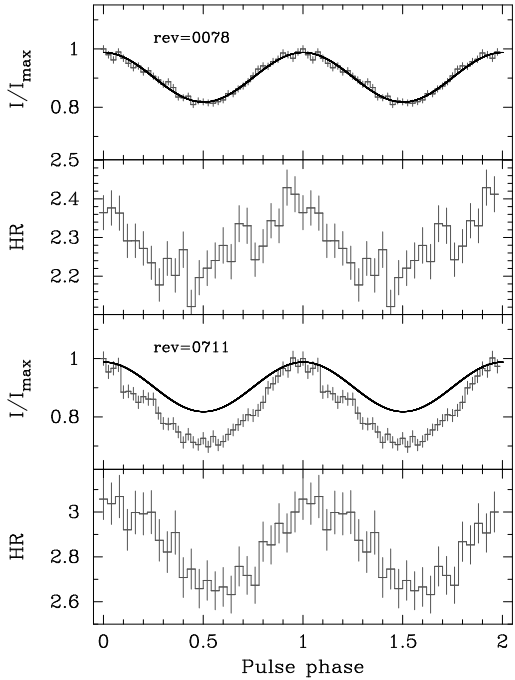,angle=270.0,width=7.0cm} \caption[hardness]{
Pulse profile of \rxj\ in the 0.1--1.2 keV band (panels 1 and 3 from
top to bottom) and hardness-ratio light curve (panels 2 and 4) for
revolutions 0078 and 0711 using EPIC/PN. The 0.1--1.2 keV light curves
are normalised to 1 at the maximum. Phase 0 is defined when the
full-band light curve reaches its maximum. The solid line is the best
sinusoidal fit to the full-band light curve of revolution 0078. The
same sinusoidal fit is overplotted on top of the full-band light curve
of the other revolution for comparison. For both observations the thin
filter was used.
}
\label{pulse}
\end{figure}	

Starting from the raw data, we first produce a list of calibrated
events. To reduce pile-up, in the next step we select only single
events as well as events that are not affected by some of the
imperfections (bad columns, hot pixels, etc.) of the CCDs. We extract
events within a $39$ arcsec circle centred on the source. We
barycenter these events using the SAS routine BARYCEN version 1.13.4,
and we then separate the events according to their energy in 3 event
lists; the bands that we use are 0.1 to 1.2 keV, 0.1 to 0.4 keV, and
0.4 to 0.8 keV, respectively.

For each observation we find the best period in the full band using an
epoch folding technique; in all cases we find a period of 8.391 s,
consistent with the value previously found for this source by
\citet{kaplan03}. We then produce folded light curves in the three
bands,  and we also compute a folded hardness-ratio light curve from
the ratio of the 0.4--0.8 keV and the 0.1--0.4 keV light curves. In
Figure \ref{pulse} we show the 0.1-1.2 keV and the hardness-ratio light
curves. For each observation we define the phase such that the maximum
of the full-band light curve occurs at phase zero; the phase of the
hardness ratio light curves is the same as for the full-band light
curves. 

The pulse profile in the 0.1--1.2 keV band, as well as the
hardness-ratio pulse profile, change from one observation to the other.
The first panel in Figure \ref{pulse} shows a sinusoidal fit to the
pulse profile during the first observation; the same sine function is
overplotted to the full-band pulse profiles obtained from the other
observations. It is apparent that the pulse profile becomes narrower
with time.

At the same time, the hardness-ratio pulse profile also changes. In the
first observation there is a clear modulation, and the hardness-ratio
profile leads the full-band light curve by $0.064 \pm 0.017$ in phase.
In the following observations the amplitude of the hardness-ratio
modulation decreases and the
phase difference between the full-band and the hardness-ratio light
curves is consistent with zero. Eventually, in the last observation the
modulation increases again, but now the hardness-ratio light curve lags
the full band-light curve by $-0.126 \pm 0.010$ in phase.

\section{Conclusions and discussion}

The \xmm\ data of \rxj\ show that the spectrum of the source
changes on a time scale of years, the first time ever that the X-ray
spectrum of an isolated neutron star, other then soft gamma-ray repeaters or 
anomalous X-ray pulsars, is seen to change.
Whereas the changes are most pronounced in the last observation,
we think that the actual change is gradual, as witnessed by a
gradual increase in the temperatures derived from the blackbody fits;
or by a gradual increase in the index of the powerlaw in the
fits with a blackbody multiplied with a power law
(Table\,\ref{temp} and Figure\,\ref{fluxed}).
The spectral changes are accompanied by an energy-dependent change in the
pulse shape; in particular the pulse phase where the spectrum is hardest
has moved with respect to the phase of maximum flux (Figure\,\ref{pulse}).

The phase (i.e.\ angle) dependent spectrum of single neutron stars 
is currently not explained. The broad absorption features
have been interpreted as a proton-cyclotron absorption
feature \citep{haberl03a}. In pulsars with a strong field
(probably stronger than the limit for \rxj) the absorption feature
may be weakened by the strong-field quantum
electrodynamics effect of vacuum resonance mode conversion
\citep{laiho03}.
The neutron star spectra have also been interpreted as
due to cyclotron-resonance scattering of the
spectrum from the surface of the neutron star, by electron-positron
pairs in the magnetosphere \citep{ruderman03}. For both
interpretations the spectrum is
likely to be angle and energy dependent, in accordance with the
variation of the X-ray spectrum (as measured by hardness ratio)
with pulse phase.

To explain the gradual, long-term variation we consider two general
possibilities: either the intrinsic spectrum of the neutron star changes,
or our view of the neutron star changes.
The intrinsic spectrum of the neutron star could change,
due to energy release deep in the
neutron star, due to a glitch for example, causing the surface to
gradually become hotter. We consider this unlikely as the explanation
for the changing spectrum of \rxj, because it does not explain the
change in pulse shape. Also this model would predict the total flux to
increase with the temperature, in contrast to what is observed. 
Another possibility, valid for Ruderman's model, would be that the
electron-positron plasma surrounding the neutron star changes.
So far, there is no specific prediction in the Ruderman model
for changes in the magnetospheric plasma on a year-long time scale.

Therefore, we suggest that the variation in the spectrum of \rxj\ is
caused by a change in the angle under which we see the emitting
region and/or the covering electron-positron plasma, 
caused by precession of the neutron star. Precession arises
when the form of the neutron star deviates from a perfect sphere and
its rotation is not around a principal axis (as reviewed by e.g.\ Link
2003).  \nocite{link} The changes in pulse form and phase of the radio
pulsar B1828$-$11 are successfully described with free precession, and
have a time scale of years \citep{stairs}, i.e.\ comparable to the
time scale on which the variations in the spectrum and pulses of \rxj\
occur. Whether a change in viewing angle can produce the observed
hardening of the spectrum, with only a modest change in overall
flux, will depend on the details of the emission model, and requires
further investigation.

To further investigate the cause of the long-term variability
it will be useful to investigate the phase-dependence of the
spectra in more detail; to obtain a reliable period derivative
of the pulses, and look for the phase changes expected from
precession; and finally to see whether the changes continue. 

\begin{acknowledgements}
The authors whish to thank Frank Haberl for a very useful
discussion on the effect of EPIC-pn observation modes and
Jan-Willem den Herder for his continuous stimulation.
\end{acknowledgements}

\bibliographystyle{aa}
\bibliography{ns}

\end{document}